\documentclass[a4paper,11pt]{article}
\usepackage{pos}
\usepackage{xcolor}
\usepackage{caption}
\usepackage{subcaption}
\usepackage{textcomp}

\usepackage{hyperref} 
\newcommand{\orcid}[1]{\href{https://orcid.org/#1}{\includegraphics[scale=0.08]{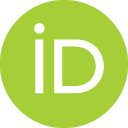}}}

\newcommand{\uRWELL}{$\mu$RWELL}

\title{$\mu$RWELL detector developments at Jefferson Lab for high luminosity experiments}

\author[a]{Kondo Gnanvo \orcid{0000-0002-5348-0664}}
\author[a]{Florian Hauenstein \orcid{0000-0002-1265-2212}}
\author[a]{Sara Liyanaarachchi \orcid{0000-0001-9136-187X}}
\author[b]{Nilanga Liyanage \orcid{0000-0002-8983-0902}}
\author[b]{Huong Nguyen \orcid{0009-0007-2525-3628}}
\author*[a]{Rafayel Paremuzyan \orcid{0009-0001-0251-4588}}
\author[a]{Stepan Stepanyan \orcid{0000-0001-7341-6734} }

\affiliation[a]{Thomas Jefferson National Accelerator Facility, 12000 Jefferson Avenue, Newport News, VA 23606,  USA}
\affiliation[b]{University of Virginia, Department Of Physics, Charlottesville VA 22903, USA}

\emailAdd{kagnanvo@jlab.org}
\emailAdd{hauenst@jlab.org}
\emailAdd{nazeer@jlab.org}
\emailAdd{nl8n@virginia.edu}
\emailAdd{htn3r@virginia.edu}
\emailAdd{rafopar@jlab.org}
\emailAdd{stepanya@jlab.org}

\abstract{
One of the future plans at Jefferson Lab is running electron scattering experiments with large acceptance detectors at luminosities
$\mathrm{>10^{37} cm^{-2}s^{-1}}$. These experiments allow the measurements of the Double Deeply Virtual Compton Scattering (DDVCS) reaction, an important physics process in the formalism of Generalized Parton Distributions, which has never been measured because of its small cross-section. The luminosity upgrade of CLAS12 or the SOLID detector makes Jefferson Lab a unique place to measure DDVCS. One of the important components of these high luminosity detectors is a tracking system that can withstand high rates of $\mathrm{\approx 1MHz/cm^{2}}$. The recently developed Micro-Resistive Well (\uRWELL) detector technology is a promising option for such a tracking detector by combining good position resolutions, low material budget with simple mechanical construction, and low production costs. In this proceeding, we will discuss recent developments and studies with \uRWELL\ detectors at Jefferson Lab for future upgrades of the CLAS12 detector to study the DDVCS reaction.}

\FullConference{%
  $\mathrm{10^{th}}$ International Conference on Quarks and Nuclear Physics (QNP2024) \\
  8-12 July 2025\\
  University of Barcelona, Barcelona, Spain
}

\begin{document}
\maketitle

\section{Introduction } 

Future directions envisioned at Jefferson Lab (JLab) include the development of large acceptance, high luminosity facilities \cite{Arrington:2021alx}. Such setups will allow measurements of small cross-section electron scattering reactions in multidimensional phase space that are impossible to measure in a reasonable time frame with currently available experimental configurations. A lot of information on the three-dimensional imaging of the quark structure of the nucleon has already been obtained from Exclusive and Semi-Inclusive DIS experiments at JLAB, and more data are expected after the completion of the approved experimental program. However, there is important information on the internal dynamics of partons that the current experiments cannot reveal. This information can be obtained by studying the Double Deeply Virtual Compton Scattering process (DDVCS) \cite{Guidal:2002kt}. Two Letter-Of-Intents (LOIs) have been submitted to the JLab Program Advisory Committee, which intend to measure DDVCS in Jefferson Labs experimental Hall B \cite{LOI:DDVCSCLAS122016} with relatively small modifications of the
existing CLAS12 detector and in Hall A with SOLID detector \cite{LOI:DDVCSSOLID2023}. The LOIs propose to run at $\ge \mathrm{10^{37}cm^{-2}s^{-2}}$ luminosity, which is $\times 100$ 
higher than the standard CLAS12 luminosity \cite{Burkert:2020akg}. 
For these detectors, high-rate capable tracking detectors are a must as at such high luminosities, the flux of particles on large acceptance detectors is very high, $\ge\mathrm{\sim 1\;MHz\;cm^{-2}}$. 
Recently developed Micro Resistive Well (\uRWELL) detectors \cite{Bencivenni:2014exa} are promising candidates for tracking detectors because of their low material budget and relatively simple design.
In this paper we will discuss currently ongoing tests with \uRWELL\ detector prototypes at JLab, which includes four small $\mathrm{10\;cm\times 10\;cm}$ prototypes
designed for very high rate ($\mathrm{> 1\; MHz/cm^{2}}$) environments, and another large prototype $\mathrm{\approx 150\;cm\times 50\;cm}$, which is not designed for very high
luminosities, but rather for assisting CLAS12 tracking for only $\times 2$ higher luminosity.


\section{ Micro Resistive Well detectors } 


The $\mu$RWELL detector is a type of Micro-Pattern Gaseous Detector (MPGD) that utilizes a single amplification stage integrated with a resistive layer. This design improves operational stability and provides effective spark protection, making it well suited for high-speed environments. Its compact design uses minimal material, making it cost-effective and easy to manufacture.

\begin{figure*}[htp]
    \centering
    
    \begin{subfigure}[b]{0.48\linewidth}
    \centering
         \includegraphics[width=0.78\linewidth]{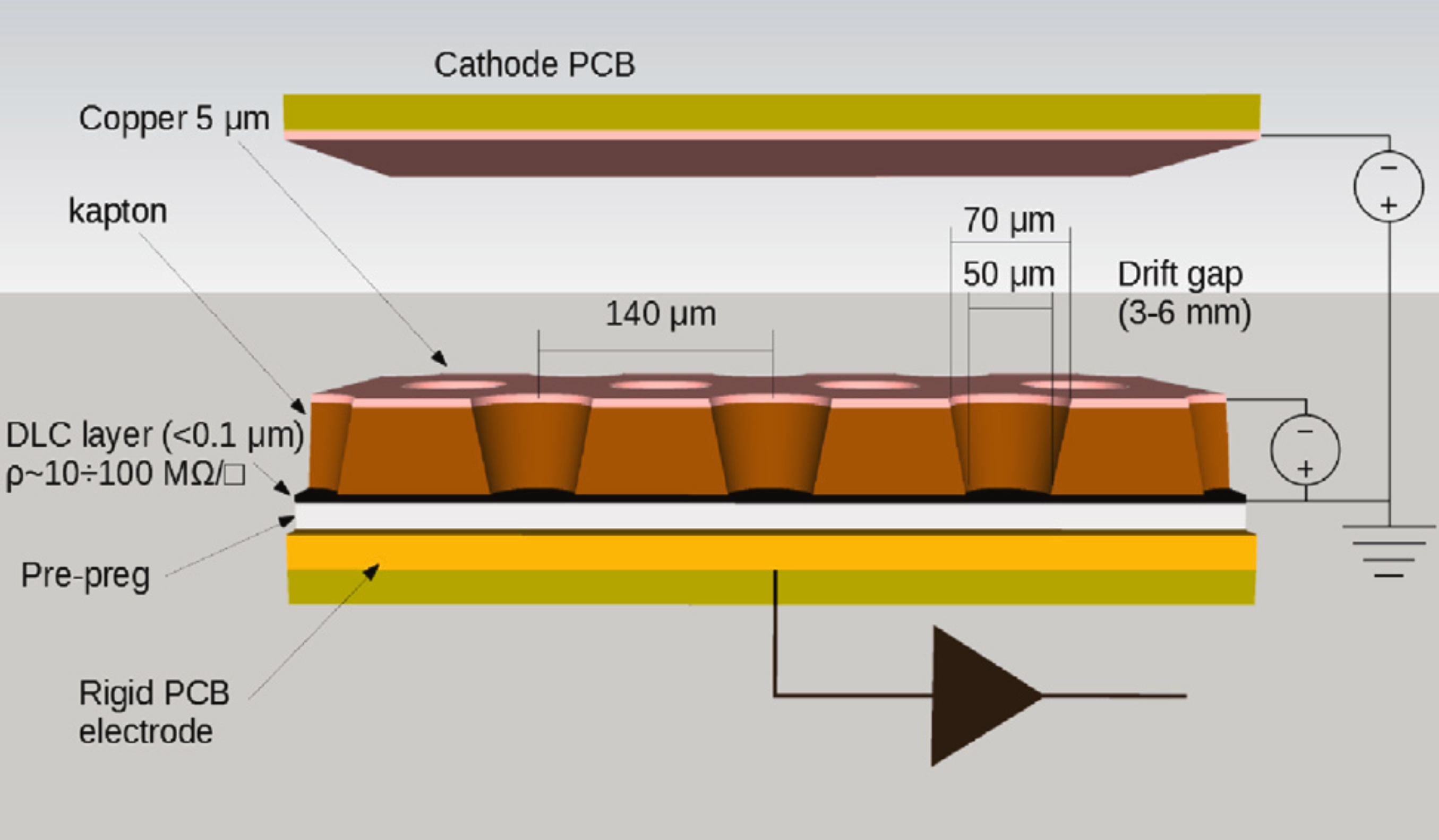}
    \caption{$\mu$-RWELL Schematic}
    \label{fig:urwell}
    \end{subfigure}
    \begin{subfigure}[b]{0.48\linewidth}
      \centering
        \includegraphics[width=0.99\linewidth]{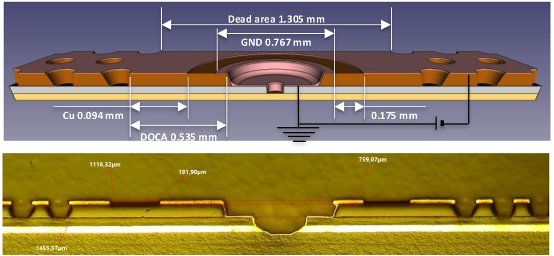} 
        \caption{Schematic of PEP-dot structure.}
        \label{fig:pepdot}
    \end{subfigure}   
    \hfill
    \caption{(Left) Schematic drawing of a $\mu$-RWELL detector \cite{Bencivenni:2024}. (Right) PEP-dot structure in a $\mu$-RWELL detector for charge evacuation \cite{pepdotbencivenni:2024}.}
    \label{fig:item1results}
\end{figure*}

The detector consists of two primary components, as shown in Fig.~\ref{fig:urwell}: the cathode and the multilayer \uRWELL\ printed circuit board (PCB). The $\mu$RWELL-PCB is constructed with a thick polyimide (apical\textregistered) foil, featuring a copper layer on the upper surface and a sputtered coating of Diamond-Like Carbon (DLC) on the bottom side. The foil is perforated with microwells, which serve as the amplification sites for the ionized electrons produced when a charged particle passes through the detector. The introduction of the resistive layer significantly suppresses the transition from a large localized discharge to a spark, allowing the detector to achieve higher gains without compromising its efficiency in high-flux environments, providing spark protection. Below the DLC are several layers of PCB to read the induced signals and ensure efficient charge collection across the detector. Some MPGDs have a capacitive sharing readout structure that enables high spatial resolution while reducing the required number of readout channels \cite{Gnanvo:2023tgf}.

The initial designs of a $\mu$RWELL detector have a Single Resistive Layer (SRL) that has a grounding line around the entire active area. The time taken for the induced charges to reach the ground, which depends on the particle's point of incidence, introduces limitations on the rate capability of the detector. During the past several years, various R\&Ds on high-rate layouts \cite{Bencivenni_2019} were tested, and recently introduced PEP (Patterning-Etching-Plating) shows promising results in high-particle flux environments. The PEP scheme is realized by creating a grounding network in the resistive stage, which is formed by etching through the top copper layer, the Kapton foil, and down to the DLC layer \cite{Bencivenni:2024}. 
The most efficient way to realize the grounding network is by small dots (PEP-dot) in a pattern structure in the detector. An example of such a structure is shown in Fig.~\ref{fig:pepdot}. It allows for an optimized grounding with the least amount of dead areas within the detector.


\section{ Experimental setup at Jefferson Lab for testing high rate detectors } \label{sec:ExpSetup}

At Jefferson Lab, we currently test various designs of \uRWELL\ detectors, which should be able to operate at particle rates of $> 1 \mathrm{MHz}/\mathrm{cm}^{2}$ utilizing the advancement in grounding schemes with PEP. The testing is part of a Laboratory Directed Research and Development (LDRD) program which promotes research on detector technologies for future experiments at Jefferson Lab.
We designed four $10\,\textrm{cm}\ \times 10\,\textrm{cm}$ \uRWELL\ prototypes with the support from the CERN MPGD workshop, which also built the detectors.
The main structure of each detector is similar. However, different aspects of each detector were varied to understand their influence on operation, efficiency and resolution at high-rates. 

The grounding schema of the DLC uses PEP-dots (Fig.~\ref{fig:pepdot}) to achieve simultaneously small dead areas and short ground lines for high-rate capabilities. Three of the four prototypes have a dot pitch of $2\,\textrm{cm}$ (hence 6$\times$6 dots in a $10\,\textrm{cm} \times 10\,\textrm{cm}$ prototype), while one prototype has a pitch of $1\,\textrm{cm}$, which doubles the dot density. This prototype will have a larger dead area, but we expect a higher rate capability than the others.
The readout structure of the prototypes has XY strips (with $800\,\mathrm{\mu m}$ pitch, hence 128 channels per readout direction) and includes capacitive sharing layers \cite{Gnanvo:2023tgf}. One of the prototypes has an additional U layer at the top of the wells with a pitch of $1600\,\mathrm{\mu m}$ to study the improvement on hit ambiguities.
Usually the \uRWELL\ well pitch is $140\,\mathrm{\mu m}$. However, one prototype has a decreased pitch of $100\,\mathrm{\mu m}$ to increase the average gain from charge deposition in the drift gap, which might be critical under high rates where the gain can change.
Table~\ref{tab:prototypeoverview} summarizes the designs of the four prototypes. Overall, pairs of prototypes have one parameter that is varied.

\begin{table}
    \centering
    \begin{tabular}{c||c|c|c|c}
        \textbf{Prototype} & \textbf{PEP dot pitch} & \textbf{Readout Type} & \textbf{Readout pitch} & \textbf{Well pitch} \\ \hline\hline
         1 & 2 cm & XY & $X=Y=800\,\mu\mathrm{m}$  & $140\,\mu\mathrm{m}$\\ \hline
         2 & 1 cm & XY & $X=Y=800\,\mu\mathrm{m}$ & $140\,\mu\mathrm{m}$\\ \hline
         3 & 2 cm & XY & $X=Y=800\,\mu\mathrm{m}$ & $100\,\mu\mathrm{m}$\\ \hline
         4 & 2 cm & XYU & $X=Y=800\,\mu\mathrm{m}, U = 1600\,\mu\mathrm{m}$ & $140\,\mu\mathrm{m}$\\ \hline
    \end{tabular}
    \caption{Design parameters for the four small $10\,\textrm{cm} \times 10\,\textrm{cm}$ \uRWELL\ prototypes which are currently tested at Jefferson Lab for high-rate capabilities. The prototypes vary in grounding dots pitch, readout, and well pitch size, and readout type.}
    \label{tab:prototypeoverview}
\end{table}\begin{figure}[!htb]
    \centering
    \includegraphics[width=0.6\linewidth]{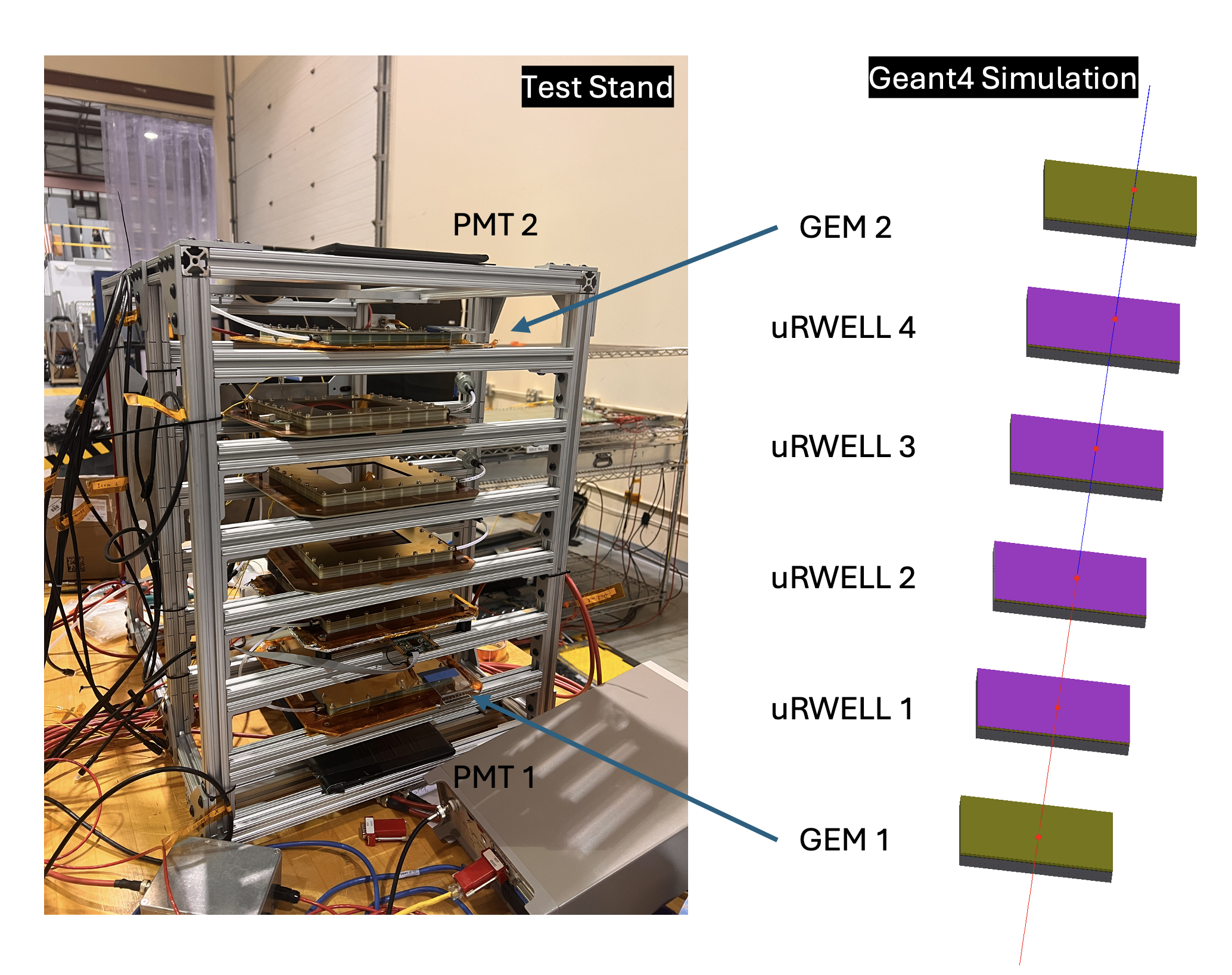}
    \caption{(Left) Experimental setup for testing of small \uRWELL\ prototypes with cosmic particles. (Right) View of setup in Geant4 simulations with a particle crossing each detector.}
    \label{fig:Teststand}
\end{figure}
The detector test stand is shown in Fig.~\ref{fig:Teststand} (left). The setup consists of two scintillation counters for triggering, two $10\,\textrm{cm} \times 10\,\textrm{cm}$ GEM detectors as reference trackers, and the 4 \uRWELL\ prototypes sandwiched between the scintillators and GEMs. The detectors are also implemented with their respective geometry and readout in Geant4 simulations. An example of a track going through the detectors is shown on the right side of Fig.~\ref{fig:Teststand}. The readout of each detector is done via APV25 front-end boards, which are connected to an SRS crate \cite{Martoiu:2011zja}. The photomultiplier tubes of the trigger counters are read out with FADC250 boards \cite{fADC:250}. The test stand is flexible and can be later rotated by any angle for test measurements in the experimental halls at Jefferson Lab. In the halls, a high-rate environment from particles is produced from interactions of multi-GeV electrons with targets.

\begin{figure}[!tb]
    \centering
    
    \begin{subfigure}{0.3\textwidth}
        \centering
        \includegraphics[width=0.95\linewidth]{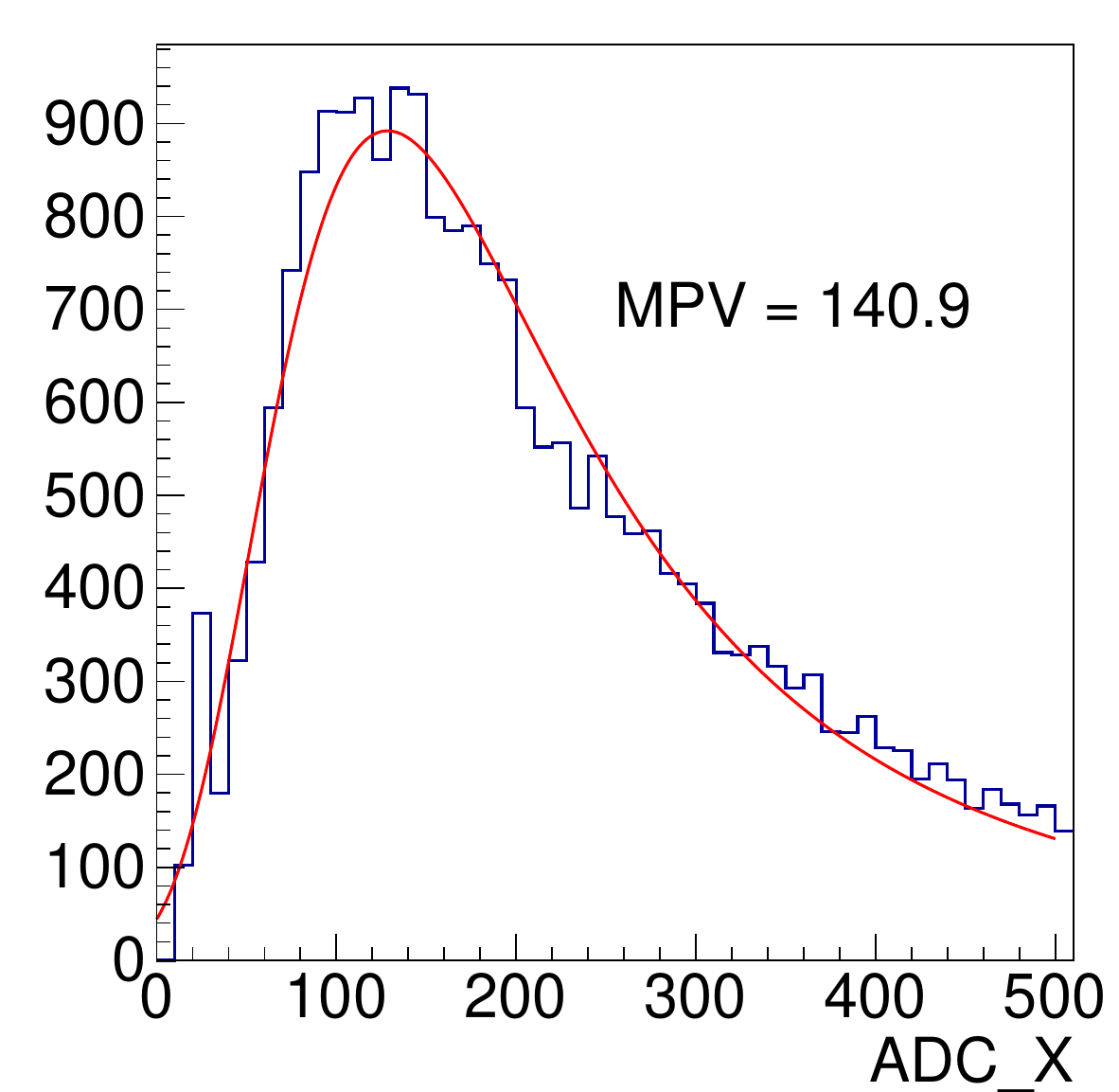} 
        \caption{ADC Distribution for X-Strips.}
        \label{fig:Item1_ADC_X}
    \end{subfigure}
    \hfill
    \begin{subfigure}{0.3\textwidth}
        \centering
        \includegraphics[width=0.95\linewidth]{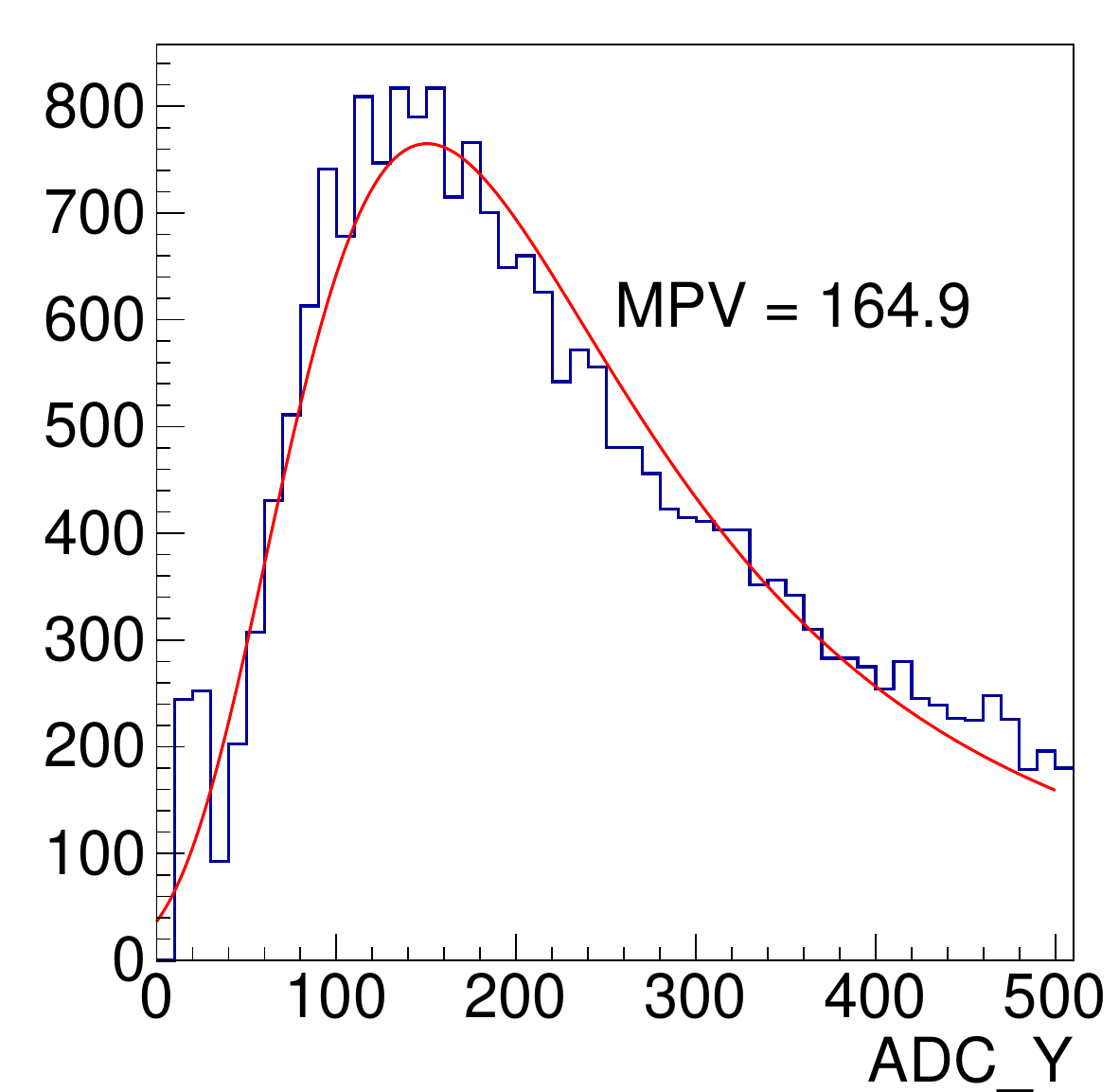} 
        \caption{ADC Distribution for Y-Strips.}
        \label{fig:Item1_ADC_Y}
    \end{subfigure}
    \hfill
    \begin{subfigure}{0.3\textwidth}
        \centering
        \includegraphics[width=0.95\linewidth]{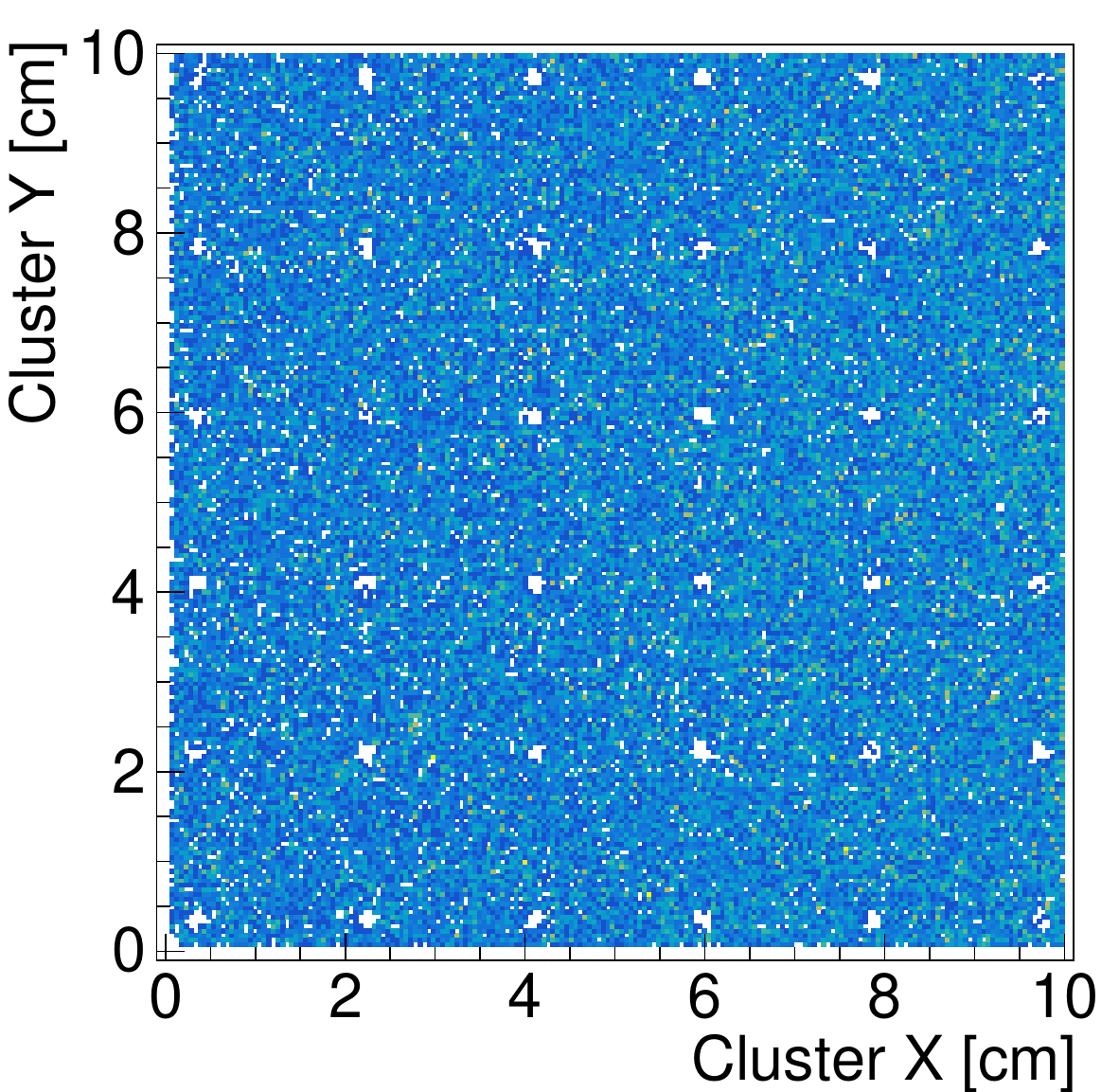} 
        \caption{ 2D hit position map.}
        \label{fig:Item1_XY}
    \end{subfigure}
    \hfill
    \caption{First results from prototype 1 with cosmic particles. ADC distributions for readout planes X (a) and Y (b). (c) 2D hit position map, which show holes from grounding dots.}
    \label{fig:item1results}
\end{figure}

The first tests of the prototypes with cosmic particles show expected results for the ADC signal in both readout layers as well as a uniform distribution across the detector surface. Preliminary plots are shown in Fig.~\ref{fig:item1results} for prototype 1. The ADC distributions (Fig.~\ref{fig:Item1_ADC_X} and \ref{fig:Item1_ADC_Y}) follow a Landau distribution as expected and are similar for both directions.
The 2-D cluster distribution (Fig~\ref{fig:Item1_XY}) has inefficient areas (white dots), which come from the PEP-dot grounding with a 2 cm spacing in both directions. 
More studies with cosmics are in progress to study the dependence of efficiency and resolutions on HV settings and different gas mixtures before high-rate beam tests.


\section{ Large prototype detector for CLAS12 luminosity upgrade } 

We also study large area \uRWELL\ detector for the CLAS12 forward tracking to support operations at twice the standard luminosity ($\mathrm{2\times 10^{35}cm^{-2}s^{-1}}$), see more 
details in \cite{LumiUpgradeTaskForceReport}.
The active area of the large prototype \uRWELL\ detector has an isosceles trapezoid shape, with a height of 50 cm, a 145 cm large base, and a 101 cm smaller base.
This is the largest \uRWELL\ detector built so far.
Unlike the small prototypes described in section \ref{sec:ExpSetup}, which are made for testing high rate capabilities of \uRWELL\ detectors,
the large-size prototype was built to study the performance of the \uRWELL\ detector with long strips and high capacitance, to understand 
how the large size impacts the overall performance of the detector in terms of efficiency, noise and the gain.
\begin{figure}[!hbt]
    \centering
    \includegraphics[width=0.99\linewidth]{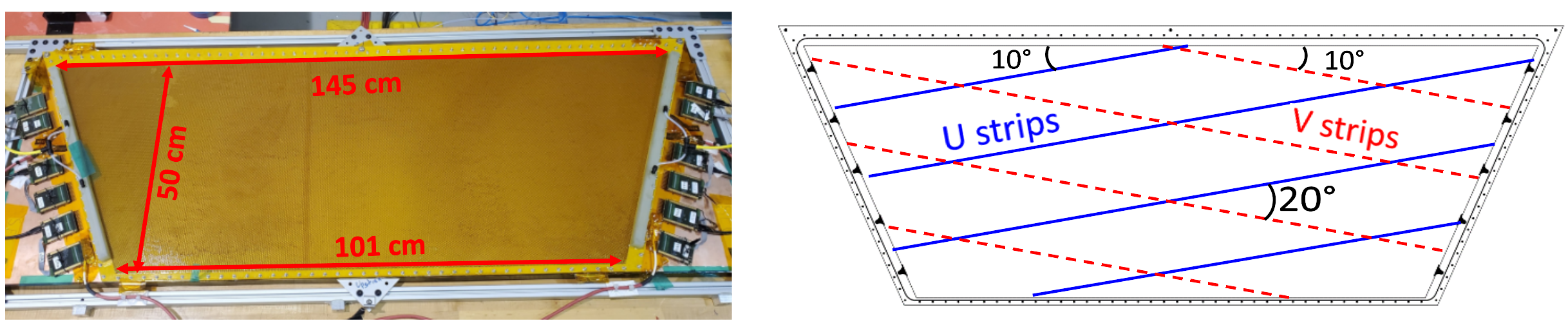}
    \caption{(Left) Photo of the large \uRWELL\ prototype detector. (Right) The schematic drawing describing the orientations of readout strips.
    U strips are represented by a blue solid line, V strips are shown by a red dashed line.
    Note: In the picture only 4 lines are shown from each strip orientation - the actual number of U(V) strips is 704(704). }
    \label{fig:BigProtoPhoto}
\end{figure}
A picture of the large \uRWELL\ prototype detector is shown in Fig.~\ref{fig:BigProtoPhoto} on the left. On the right side, the 
orientations of the readout strips are shown. There are two layers of readout strips defined as U and V, oriented $\mathrm{\pm 10^{\circ}}$ with respect to the base of the detector. The strip pitch is 1 mm for both U and V strips. 
The first readout layer is the U layer, followed by the V layer.
\begin{figure}[!htb]
    \centering
    \begin{subfigure}[n]{0.49\textwidth}
        \centering
        \includegraphics[width=\linewidth]{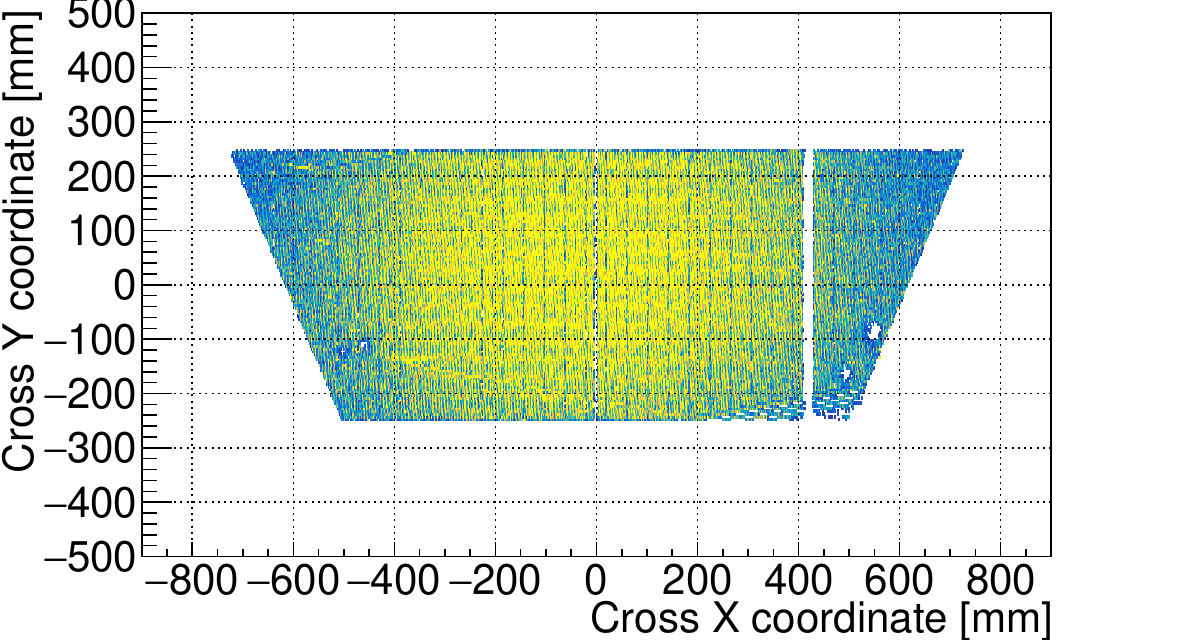}
        \caption{``Y vs X'' hit distribution.}
        \label{fig:BigProto_Cross_YXc_Max1}
    \end{subfigure}
\hfill
    \begin{subfigure}[n]{0.49\textwidth}
        \centering
        \includegraphics[width=\linewidth]{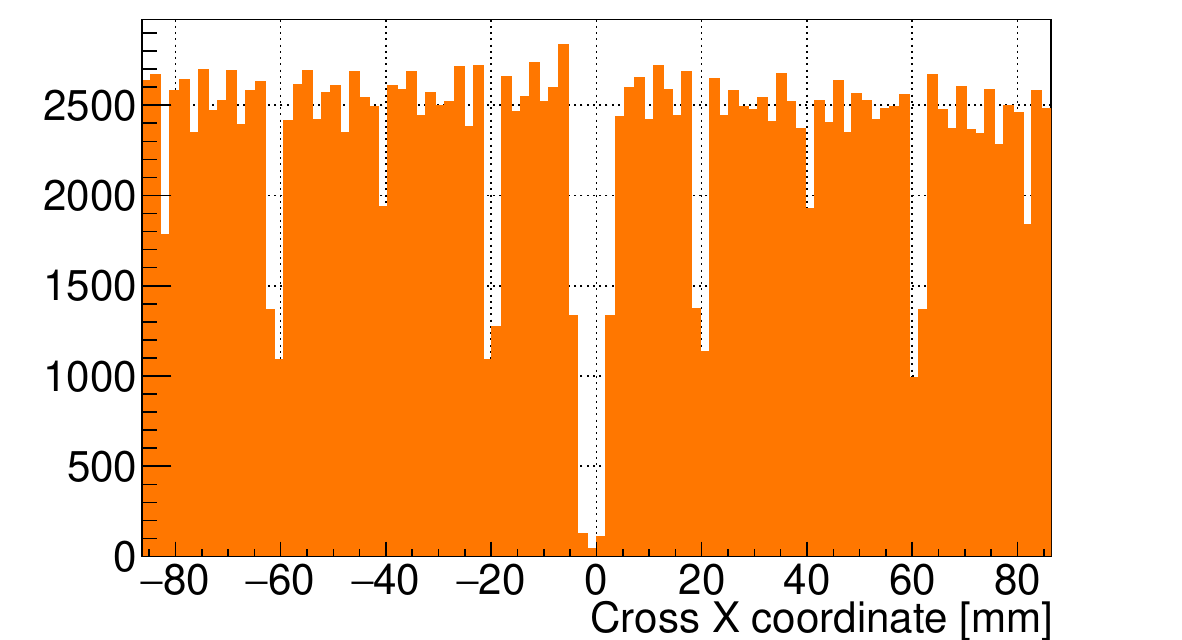}
        \caption{X hit distribution zoomed in the range -85 mm to 85 mm.}
        \label{fig:BigProto_Cross_Xc_Max1}
    \end{subfigure}

    \caption{Distribution of hit coordinates on the detector surface from the cosmic data. (Left) ``Y vs X'' hit distribution. (Right)
    X hit distribution in the -85 mm to 85 mm range. }
    \label{fig:BigProtoOccupancies}
\end{figure}
The large prototype was tested with cosmic particles. The hit occupancy from these tests is shown in
Fig.~\ref{fig:BigProtoOccupancies}. Several distinct features and patterns are noticeable:
\begin{enumerate}
    \item 
 The detector has a $\sim 20$ mm wide region at $\mathrm{x}$ around $\mathrm{\in(410\; cm - 430\; cm}$, with essentially no occupancy. This region was present from the beginning of the tests and was found to be an HV issue in that section of the $\mu$RWELL.
\item There are four round-shaped low occupancy spots, two around ($x,y$) at ($-500\,\textrm{mm}$,$-120\,\textrm{mm}$), and two around ($+500\,\textrm{mm}$,$-100\,\textrm{mm}$). These inefficient regions appeared after the detector was opened to replace the cathode and are most likely due to dust particles trapped inside the wells. The fact
that the detector developed a resistive-like current (i.e., current proportional to the applied voltage, around 10 $\mathrm{\mu A}$ at 480 V) after the cathode exchange supports the dust trapped in the wells hypothesis. Despite the dust in the detector, it is still operational with only a relatively small inefficient region around the dust particle.
\item There is a periodic structure in X (more apparent in Fig.~\ref{fig:BigProto_Cross_Xc_Max1}) with a period of 20~mm. This structure corresponds to HV sections of the detector, and dips in the distribution reflect the gaps between two HV sections.
\end{enumerate}


\begin{figure}[!htb]
    \centering
    \begin{subfigure}[b]{0.45\textwidth}
        \centering        
        \includegraphics[width=0.85\linewidth]{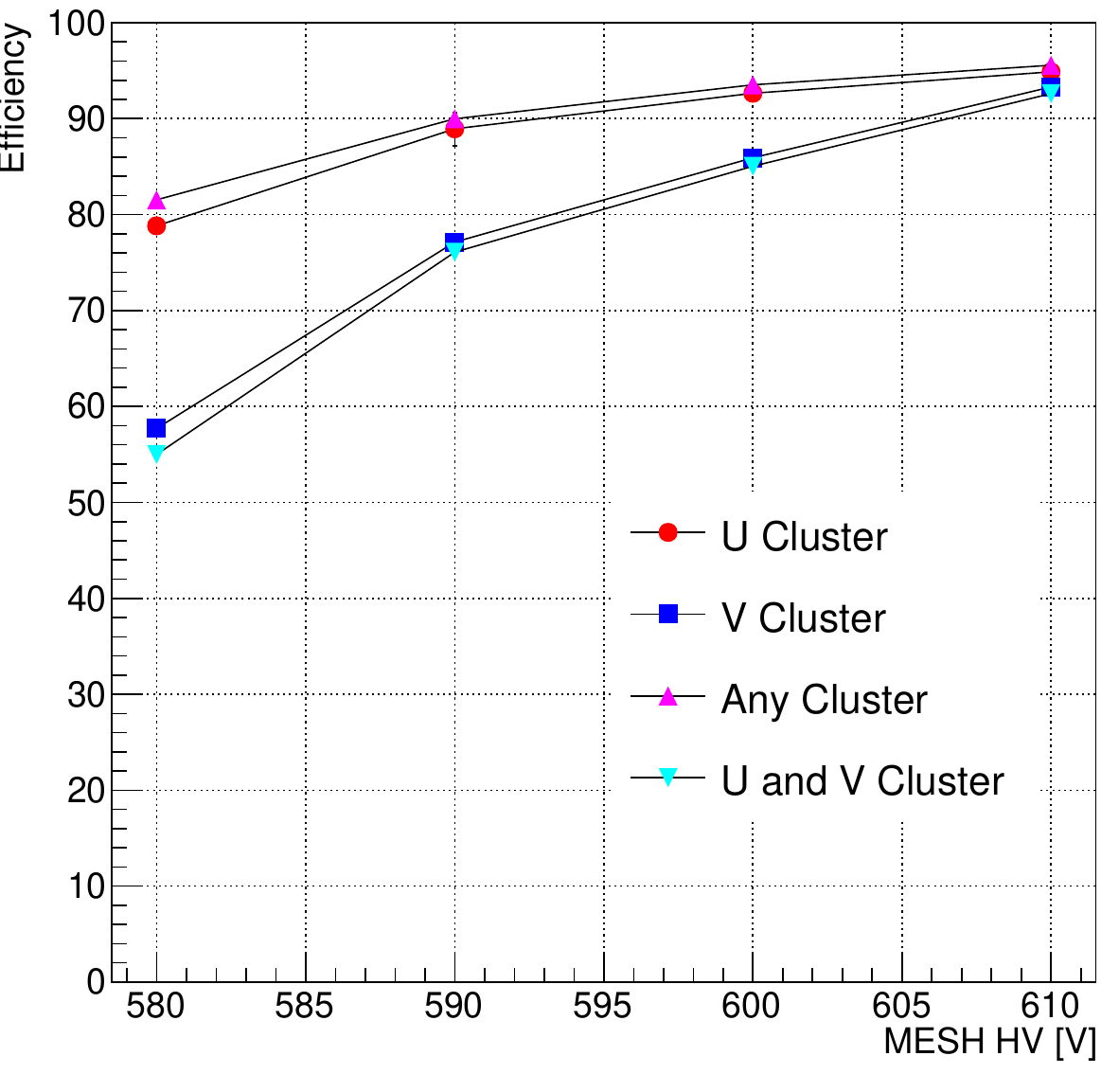}
        \caption{Efficiency with an $\textrm{Ar:CO}_2$ gas 80:20 mixture.}
        \label{fig:ArCO2_BigProtoEff}
    \end{subfigure}
\hfill
    \begin{subfigure}[b]{0.45\textwidth}
        \centering        
        \includegraphics[width=0.85\linewidth]{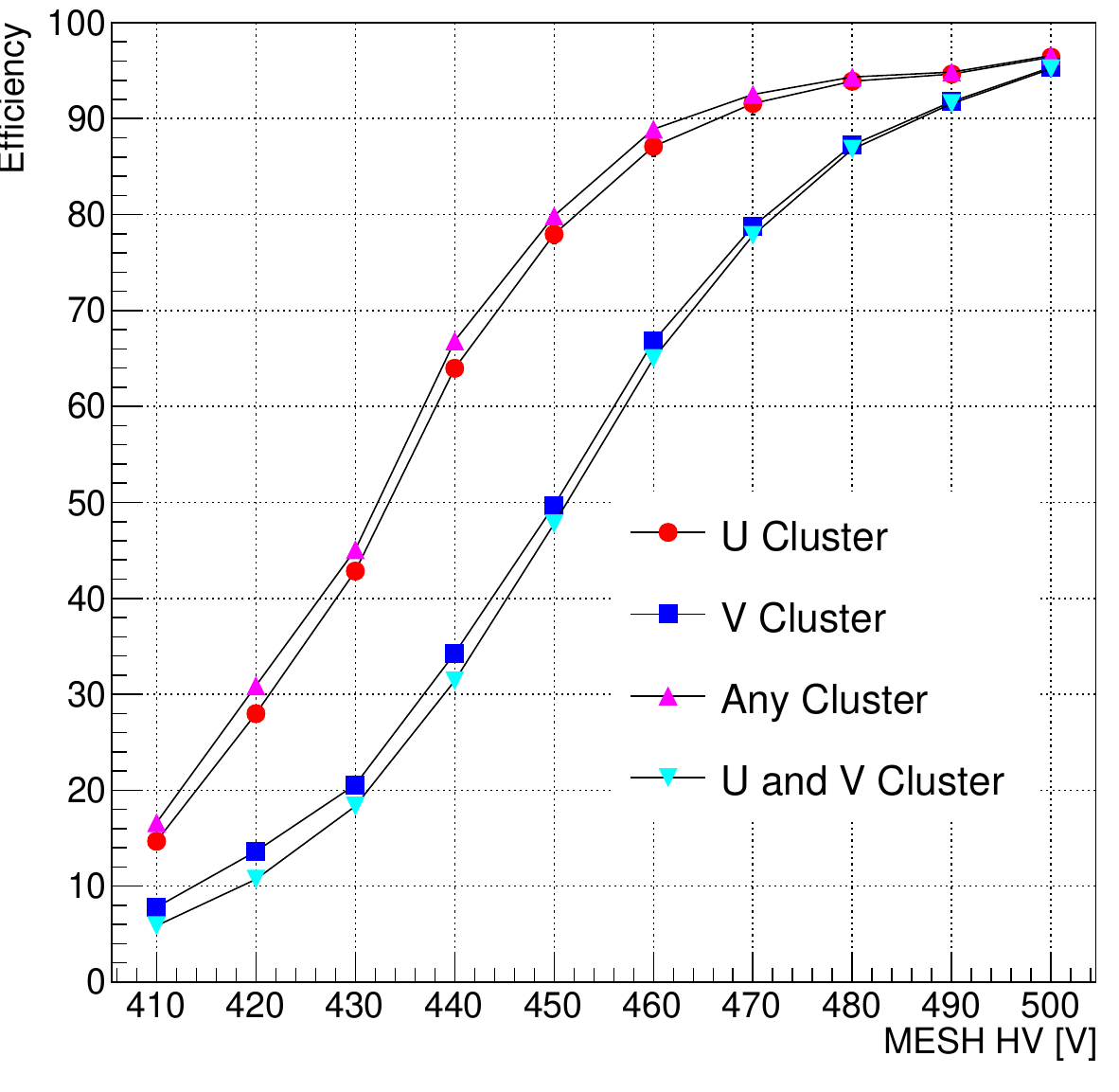}
        \caption{Efficiency with an $\textrm{Ar:C}_{4}\textrm{H}_{10}$ gas 90:10 mixture.}
        \label{fig:ArIso_BigProtoEff}
    \end{subfigure}

    \caption{Efficiency as a function of HV for $\textrm{Ar:CO}_2$ gas 80:20 mixture (left) and for $\textrm{Ar:C}_{4}\textrm{H}_{10}$  gas 90:10 mixture (right). Red circle markers represent the U layer efficiency,  blue squares represent the V layer efficiency, 
    upright triangles in pink is the efficiency of a hit in any layer, and inverted triangles in cyan represent
    the efficiency of a hit to be present in both, U and V layers.}
    \label{fig:BigProto_Efficiency}
\end{figure}

We have also studied the efficiency of the detector with two different gas mixtures: $\textrm{Ar:CO}_2$ (80:20) and with $\textrm{Ar:C}_{4}\textrm{H}_{10}$ (aka Isobutane) (90:10). The
efficiencies of U and V strips are studied separately and together as a cluster. We observe that for both gas mixtures,
the efficiency of U strips is higher than the efficiency of V strips, see Fig.~\ref{fig:BigProto_Efficiency}.
This is not surprising as the U layer is closer to the amplification stage. 
For the drift field (field between the cathode and the \uRWELL\ MESH), 
we found that above 2500 V/cm, we do not get any improvement in gain or efficiency.
Therefore, we kept the drift field at 2500 V/cm for most of our studies.
We started the test with the $\textrm{Ar:CO}_2$ mixture, as it is widely used for GEM detectors as a standard gas and also in previous measurements of \uRWELL\ detectors  \cite{Gnanvo:2023tgf}.
Eventually, we were able to reach > 90\% efficiency for both U and V strips at 610 V. However, the detector was very unstable, and we could observe frequent spikes in the
leakage current. Afterward, we switched to the $\textrm{Ar:C}_{4}\textrm{H}_{10}$ mixture, and we observed that the detector is much more stable with >90\% efficiency at 490 V.
Pushing the HV to 500V, some discharges were observed on the leakage current, however the rate was much smaller (1-2 per hour) in comparison to the  $\textrm{Ar:CO}_2$ gas.

\section{Summary and outlook }
Despite some issues in the performance of the large prototype \uRWELL\ detector, we were able to reach high efficiency ($\mathrm{>90\%}$)
using two different gas mixtures: Ar/CO2 (80/20) and Ar/Isobutane (90/10). However, the detector was not quite stable with the Ar/CO2 gas mixture. 
Under the Ar/Isobutane mixture, a high efficiency is reached without significantly impacting the detector stability.
Incidental observation during tests suggests that even with dust particles deposited on the amplification wells, the detector can still be operational with the cost of having small dead areas around the dust particles.
For the large \uRWELL\ detector, we plan to build a new version as a pair of 1D detectors with the same active area.
The orientation of the strips will be $\mathrm{\pm 10^{\circ}}$ with respect to the base of the detector. We expect to get higher gains and, hence, larger headroom on HV for operations with beam.

Initial quality checks of the small prototype detectors did not reveal any issues. The plan for the next several months is to study the stability and efficiency of the detectors with different gas mixtures and HV settings on the cosmic test stand and then move the detectors to an experimental hall in a
high-rate environment in early 2025 to test the high-rate capability. 

\section{Acknowledgments} 
The authors are thankful to Bob Miller and Christopher Guthrie for the design of the support structures for both small and large-size prototypes,
to Rui De Oliveira and Bertrand Mehl for the design and construction of the \uRWELL\ foils and readout,
to Sergey Boyarinov and Xinzhan Bay for supporting data acquisition
and Raffaella De Vita and Maurizio Ungaro for their support in the GEANT4 simulation.

This material is based upon work supported by the U.S. Department of Energy, Office of Science, Office of Nuclear Physics under contract DE-AC05-06OR23177 

\end{document}